\documentclass{aa}  

\usepackage{amsmath}
\usepackage{txfonts} 
\usepackage{hyperref}
\usepackage{graphicx} 
\usepackage{caption}
\usepackage{subcaption}
\usepackage[dvipsnames]{xcolor}
\usepackage{booktabs}

\newcommand{\comment}[1]{} 

\begin{document}

\title{Numerical modelling of recombination-driven stellar winds}


\author{
Bestin James\corrauth{bestin@physics.muni.cz}
\and 
Jiří Krtička
}


\institute{Department of Theoretical Physics and Astrophysics, Faculty of Science, Masaryk University, 611 37 Brno, Czech Republic
}

\date{}

  \abstract
    {Low-temperature stars have a significant population of neutral hydrogen
    increasing with height in their atmospheres. This suggests
    eventual recombination as we move away from their deep photospheric layers. 
    Recombination can potentially drive stellar winds, particularly in cool and evolved stars.}
   {We aim to verify the astrophysical possibility of stellar winds driven by recombination. We also aim to constitute an initial model for this type wind driving.} 
   {We examine the possibility of recombination-driven outflows across the Hertzsprung–Russell (HR)
   diagram by comparing the gravitational potential energy per atom to the hydrogen ionization energy. Then we start from a simple analytical model with the addition of an extra heating term to the gas in the classical Parker wind model. We model the heating term, analogous to the heat released by recombination, and examine how it can drive an outflow. We implement the model in the numerical code CASTRO and check the hydrodynamical stability of the solution.}
   {The added heating drives a steady transonic outflow in our models, when tested with the parameters of an asymptotic giant branch (AGB) star. The properties of the heating function affect the solution and the location of the sonic point depends on the location of the heating function itself. 
   Even though the derived hydrodynamical equations are independent of the density, we determine the processes that set the mass-loss rate. By comparing the heating from collisional recombination and cooling due to radiative recombination, we show that relatively high densities are required to release the recombination energy as heat, implying mass-loss rates of the order of $0.1\,M_\odot\,\text{yr}^{-1}$.}
   {Our comparison of energies across the HR diagram suggests a possibility of recombination-driven winds in 
   evolved red giants and AGB stars. The main deciding factor for wind launching is the radial distribution of heat released by recombination in a star's atmosphere. This can in turn affect the mass-loss rate.} 


\maketitle
\nolinenumbers

\section{Introduction}\label{sec:intro}

A stellar wind is a continuous outflow from stellar 
photospheres into the interstellar medium. The first hydrodynamical 
models of stellar wind driven by thermal expansion of corona were 
computed by \citet{Parker_1958ApJ...128..664P}. The models of 
coronal wind were subsequently refined by including magnetic 
fields and wave heating \citep{Weber_1967ApJ...148..217W,Cranmer_2007ApJS..171..520C,Sakaue_2020ApJ...900..120S}.

Models describing wind driving by other effects than the thermal 
expansion typically build on Parker's model of the stellar wind. 
Asymptotic giant branch stars show pulsations that are able to 
lift photospheric matter to such distances that the metals can
condensate to dust grains. Dust particles further drive the wind 
to the interstellar medium \citep{Hoyle_1962MNRAS.124..417H,Gilman_1972ApJ...178..423G,Bowen_1988ApJ...329..299B}. Successful modelling of dust-driven stellar 
wind therefore requires a combination of the description of
stellar pulsations and radiative acceleration due to dust particles
\citep{Woitke_2006A&A...452..537W,Freytag_2023A&A...669A.155F}.

Stellar winds of hot stars are accelerated by the radiative force
originating either in absorption or scattering of stellar
radiation by line transitions of heavy elements, and Thomson 
scattering on free electrons \citep{lusol,cak}. Detailed modelling
of such winds requires global (unified) wind models that 
consistently describe not only the driving of the wind by the
photospheric radiation, but also the feedback of the
stellar winds \citep{grahamz,cmfkont,powrdyn,sundyn}.

When studying line-driven winds of A supergiants, \citet{acmfkont} noted a
region of hydrogen recombination in the subsonic parts of the wind.
The energy released as a result of hydrogen recombination can
assist in wind driving; however, the dynamical consequences of 
such a layer remain unclear. A similar effect appears in the 
asymptotic giant branch stars, where it causes 
instability of evolutionary models 
\citep{Wagenhuber_1994A&A...290..807W}. This instability
can be avoided by dumping the released hydrogen recombination
energy \citep{10.1093/mnras/stae1387}, but it can have evolutionary
consequences when the envelope is released \citep{Leon_Pl_Nebulae_1967AJ.....72Q.813L,Paczynski_1968AcA....18..255P,Han_1994MNRAS.270..121H}. Recombination has also been initially
proposed to play an important role in the winds of early-type
stars \citep{Waldron1984ApJ...282..256W}. These problems
motivate further investigation
of the implications of hydrogen recombination for wind dynamics.

Stellar winds are frequently modelled using stationary models, because the time-independent approach is typically 
easier to handle than time-dependent modelling. Therefore, 
modelling of stationary recombination-driven winds can provide
a better understanding of other situations where the hydrogen recombination could be important, such as common-envelope phase of
binary evolution. The processes that drive ejection of the common
envelope are not very well understood 
\citep{Ivanova_2013A&ARv..21...59I}, and the  hydrogen recombination is considered to be one of the possible 
mechanisms \citep[e.g.,][]{Ivanova_2015MNRAS.447.2181I}. However,
the role of other mechanisms
that may carry out the released 
energy, such as radiative transport, has been strongly debated 
\citep{Griechener_2018MNRAS.478.1818G,Soker_2018ApJ...863L..14S}.
In this case, the stationary models can help to characterise the fraction of released energy carried out by radiation.

To understand the role of hydrogen recombination, we provide models of recombination-driven stellar winds.
We first determine in which 
stars the recombination energy can lead to outflows. We 
continue with a description of stationary recombination-driven wind
models, which are further tested using hydrodynamical simulations.
We conclude with a discussion of the wind mass-loss rate. When 
finishing our analysis, we noted two additional papers dealing
with a similar topic, but using a different methodology \citep{Strusberg_2026arXiv260621624S,Yang_2026arXiv260619422Y}.

\section{Possibility of recombination-driven winds across the HR diagram}\label{sec:hrdiag-recombination}

We are interested in stars in which the available ionization energy per hydrogen atom in their atmospheres exceeds the gravitational potential energy. Fulfilment of this condition can possibly launch an outflow driven by hydrogen recombination. As noted in the introduction, there were already attempts to explain the coronae of early-type stars with recombination driven stellar wind models \citep{Waldron1984ApJ...282..256W}. It was also suggested a while ago that planetary nebulae are formed from dynamically unstable extreme red giants \citep{Leon_Pl_Nebulae_1967AJ.....72Q.813L}. As the recent work by \citet{acmfkont} noted hydrogen recombination in A-supergiants, the physical mechanism seems to be relevant in at least certain stars. 

For stellar winds to be driven by hydrogen recombination, the gravitational potential energy for the hydrogen atoms, given by $E_\text{G} = \frac{m_\text{H} MG}{R_*}$, should be comparable to energy from the recombination $I_\text{H} = 13.6$ eV.  
Expressing the stellar mass $M$ and radius $R_*$ in solar units gives a condition under which the gravitational potential is lower than the hydrogen ionization potential
\begin{equation}
\frac{m_\text{H} MG}{R_*}<I_\text{H}, \qquad 1990\,\text{eV}\,
\left(\frac{M}{M_\odot}\right)\left(\frac{R}{R_\odot}\right)^{-1}
<13.6\,\text{eV}.
\end{equation}
Stars with a low mass to radius ratio would have lower gravitational binding energies in the range comparable to hydrogen ionization energy. 
For the initial comparison, we can estimate the gravitational binding energy per hydrogen atoms using some typical stellar values. For our Sun, $E_\text{G} = 1990~\text{eV}$, which shows that the effect of recombination is negligible. But when we move on to evolved stars like extreme 
red giants or asymptotic giant branch (AGB) stars, this changes. For an AGB star with a mass of $2 M_\odot$ and a radius of $200 R_\odot$, $E_\text{G} = 19.9~\text{eV}$, which is in the order of $I_\text{H} = 13.6~\text{eV}$. So, we are mostly interested in giants or supergiants with a 
low enough surface gravity. To test this idea further, we probe stellar
evolutionary tracks in the Hertzsprung–Russell (HR) diagram across different initial masses.

\begin{figure}
    \centering
    \includegraphics[width=\linewidth]
    {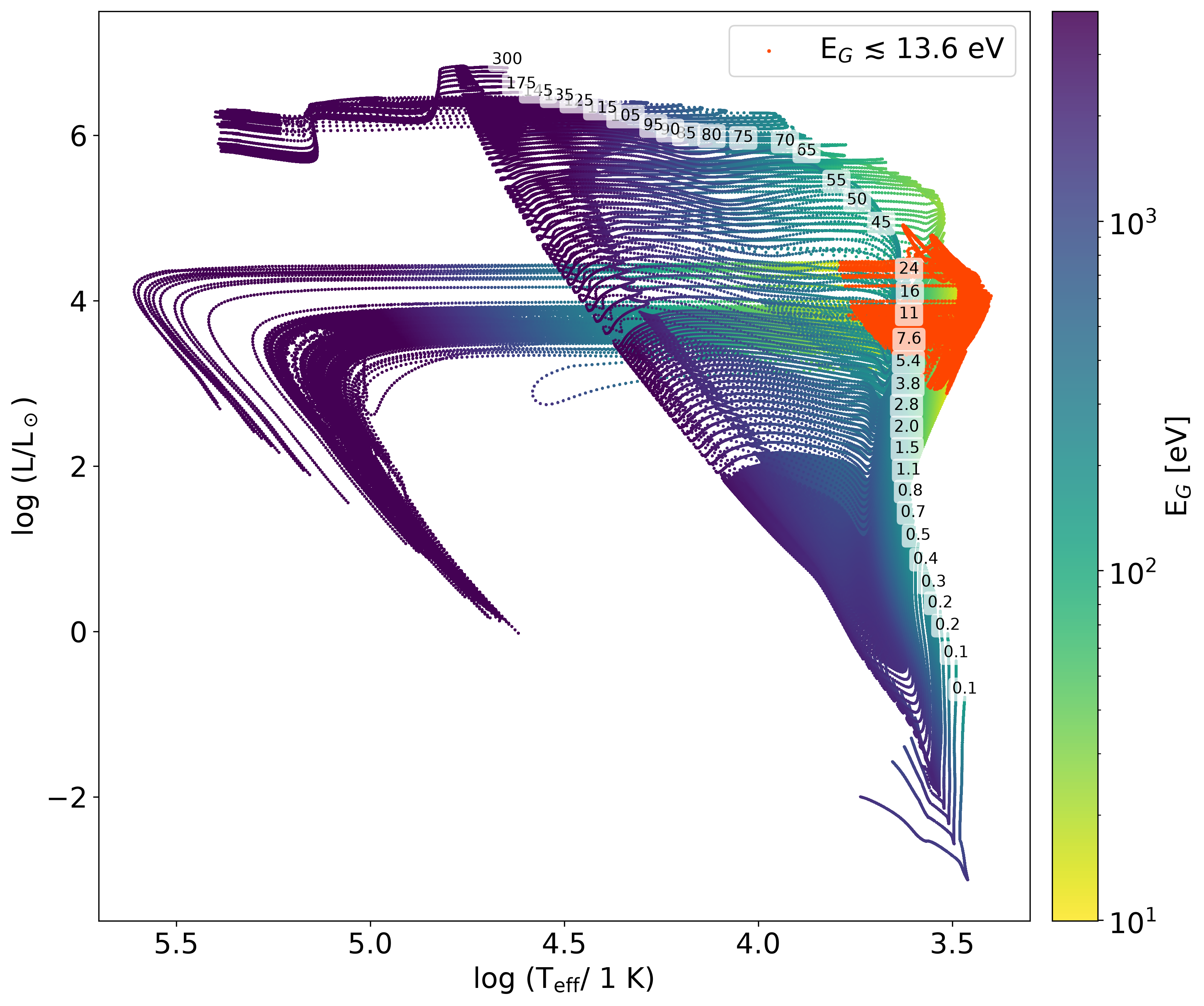}
    \caption{HR diagram showing 110 non-rotating stellar evolutionary tracks from the
    MIST database (v1.2, [Fe/H] = 0.0, $v/v_\text{crit}$ = 0.0).
    The initial masses of the tracks (in \(\text{M}_\odot\)) at their PMS
    stage are noted on selected curves. The colour map shows the gravitational binding energy per hydrogen atom ($E_\text{G}$) for each point on an evolutionary track. Stars with \(E_\text{G}\ \leq 13.6\ \mathrm{eV}\) are marked in red colour.}
    
    
    \label{fig:hr_diagram_marked}
\end{figure}

In Fig.~\ref{fig:hr_diagram_marked},
we plot an HR diagram from 110 MIST \citep[MESA isochrones and stellar tracks,][]{MIST0_2016ApJS..222....8D, MIST1_2016ApJS..222....8D, MIST_2016ApJ...823..102C}
non-rotating stellar evolutionary tracks and colour code each evolutionary point according to the gravitational binding energy per hydrogen atom in the star's atmosphere. The tracks start at the pre-main sequence (PMS) stage and evolve further. As stars reach either the red giant or AGB phase, the gravitational binding energy per hydrogen atom in their atmospheres is notably in the range of hydrogen ionization energy. These are visible as green-yellow in coloured points on the plot. In certain stars, the gravitational binding energy per hydrogen atom falls even below the hydrogen ionization energy ($E_\text{G} \leq I_\text{H}$) and these are marked in red colour. 
From the models in Fig.~\ref{fig:hr_diagram_marked}, we identify 29252 evolutionary points matching this criteria, with their masses ranging from 0.513 to 4.271 $\mathrm{M_\odot}$ and radii ranging from 76.2 to 656.6 $\mathrm{R_\odot}$. In many of these stars, the available gravitational biding energy per atom is less than half of the hydrogen ionization energy.
In these stars, the effect of hydrogen recombination is understandably relevant in driving stellar winds. We will probe this further in the following sections.

\section{Models of recombination-driven winds}
\label{sec3:model}

\subsection{Basic 
stationary
model}\label{subsec:anal_model}
We start with the usual initial assumptions as in the Parker's classical model for solar wind \citep{Parker_1958ApJ...128..664P}. Let us assume a steady, spherically symmetric, radial flow around a star of mass $M$. 
For our model, we assume the flow comprises a monoatomic ideal gas with an additional ``heating potential"
due to recombination $q(r)$ in its specific internal energy, 
\begin{equation}
    e(r) = \frac{p}{(\gamma -1)\rho} + q(r), 
\end{equation}
where  $p$ is the gas pressure, $\rho$ its density and $\gamma$ its adiabatic index. 

From mass conservation, 
\begin{equation}\label{eq:continuity1}
    \frac{1}{r^2} \frac{d}{dr} (r^2 \rho v) = 0, 
\end{equation} and we get $r^2 \rho v = 
\text{const.}
= \dot{M}/(4\pi)$, or 
\begin{equation}
    \rho(r) = \frac{\dot{M}}{4\pi r^2 v(r)},
\end{equation}
where $v$ is the gas velocity and $\dot{M}$ is the mass-loss rate. 

From momentum conservation, we write
\begin{equation}\label{eq:momentum1}
    \rho v \frac{dv}{dr} = -\frac{dp}{dr} - \rho\frac{GM}{r^2}.
\end{equation}
The energy equation, neglecting the radiative effects, the viscous and conductive terms can be written as
\begin{equation}\label{eq:energy1}
    \frac{d}{dr} \left[ \frac{v^2}{2} + h + q(r) - \frac{GM}{r}\right] = 0,
\end{equation}
where $h$ is the specific enthalpy defined as $h = e + \frac{p}{\rho} = \frac{\gamma}{\gamma-1}\frac{p}{\rho}$, for an ideal gas.

So we have the energy equation as 
\begin{equation}
    \frac{d}{dr} \left[ \frac{v^2}{2} + \frac{\gamma}{\gamma-1} \frac{p}{\rho} + q(r) - \frac{GM}{r}\right] = 0.
\end{equation}
From this, we can write our Bernoulli integral as
\begin{equation}\label{eq:energy2}
    \frac{v^2}{2} + \frac{\gamma}{\gamma-1} \frac{p}{\rho} + q(r) - \frac{GM}{r} = B,
\end{equation}
where $B$ is a Bernoulli constant, which can be determined by the physical conditions of the flow at its base. 

For a monoatomic ideal gas, 
\begin{equation}
     p = \rho \frac{k_B T}{\mu m_\text{H}},
\end{equation}
where $k_B$ is the Boltzmann's constant, $T$ is the gas temperature, $m_\text{H}$ is the mass of a proton, and $\mu$ is the mean molecular weight in units of proton mass.
Here we can also recall the definitions of isothermal and adiabatic sound speeds $a$ and $c_s$, respectively, given by
\begin{align}
    a^2 &= \frac{k_B T}{\mu m_\text{H}},\\
    c_s^2 &= \gamma a^2 = \gamma \frac{p}{\rho}.
\end{align}

Putting this into the energy equation (Eq. \ref{eq:energy2}), we get, 
\begin{equation}
\label{eq:bernoulli}
    \frac{v^2}{2} + \frac{\gamma k_B T}{(\gamma -1) \mu m_\text{H}} + q(r) - \frac{GM}{r} = B.    
\end{equation}
By rearranging, we get an expression for temperature as,
\begin{equation} \label{eq:temperature1}
    T(r) = \frac{(\gamma -1) \mu m_\text{H}}{ \gamma k_B} \left[ B - q(r) - \frac{1}{2}v(r)^2 + \frac{GM}{r}\right].
\end{equation}
From this, we can also write an explicit expression for the adiabatic sound speed as
\begin{equation}\label{eq:c_s^2__1}
    c_s^2 = \frac{\gamma k_B T}{\mu m_\text{H}} = (\gamma -1) \left[ B -q(r) - \frac{v^2}{2} + \frac{GM}{r}\right].
\end{equation}

Now we try to arrive at a differential equation for velocity. We start again from the momentum equation (Eq. \ref{eq:momentum1}).
For a polytropic gas, we have, 
\begin{equation} \label{eq:dp/dr}
    \frac{dp}{dr} = \frac{c_s^2}{\gamma}\frac{d \rho}{dr} + \frac{\rho}{\gamma}\frac{dc_s^2}{dr}.
\end{equation}
Substituting this into the momentum equation (Eq. \ref{eq:momentum1}), we get,
\begin{equation}\label{eq:vel1}
    v\frac{dv}{dr} = -\frac{1}{\rho} \frac{c_s^2}{\gamma} \frac{d\rho}{dr} - \frac{1}{\gamma} \frac{dc_s^2}{dr} - \frac{GM}{r^2}.
\end{equation}
From the continuity equation (Eq. \ref{eq:continuity1}), we can write,
\begin{equation}
    \frac{1}{\rho} \frac{d\rho}{dr} = -\frac{1}{v} \frac{dv}{dr} - \frac{2}{r}.
\end{equation}
Putting this into 
Eq.~\eqref{eq:vel1},
we get,
\begin{equation}
    v\frac{dv}{dr} = \frac{c_s^2}{\gamma} \frac{2}{r} + \frac{c_s^2}{\gamma} \frac{1}{v} \frac{dv}{dr} - \frac{1}{\gamma} \frac{dc_s^2}{dr} -\frac{GM}{r^2}, 
\end{equation}
or
\begin{equation}\label{eq:vel_1a}
    \left( v - \frac{c_s^2}{\gamma v} \right) \frac{dv}{dr} = \frac{c_s^2}{\gamma} \frac{2}{r} - \frac{1}{\gamma} \frac{dc_s^2}{dr} - \frac{GM}{r^2}.
\end{equation}

Using 
Eq.~\eqref{eq:c_s^2__1} for the adiabatic sound speed, we can write its derivative as,

\begin{equation}\label{eq:dcs2/dr}
     \frac{dc_s^2}{dr} = - (\gamma - 1)\left[ \frac{dq}{dr} + v \frac{dv}{dr} + \frac{GM}{r^2}\right],   
\end{equation}
which can then be substituted to the velocity equation (Eq. \ref{eq:vel_1a}).

\subsection{The added heating function}\label{subsec:heating_func}

We are interested in stellar winds driven by hydrogen recombination. To better understand this problem, we can assume a heating potential of the form $(I_\text{H}/m_\text{H}) X_\text{H}$,
where $I_\text{H} = 13.6~\mathrm{eV}$ and $X_H$ is the hydrogen ionization fraction.
We introduced two trial functions with different radial variations of fraction of ionized hydrogen.

\begin{figure}
[t]
    \centering
    \includegraphics[width=0.8\linewidth]{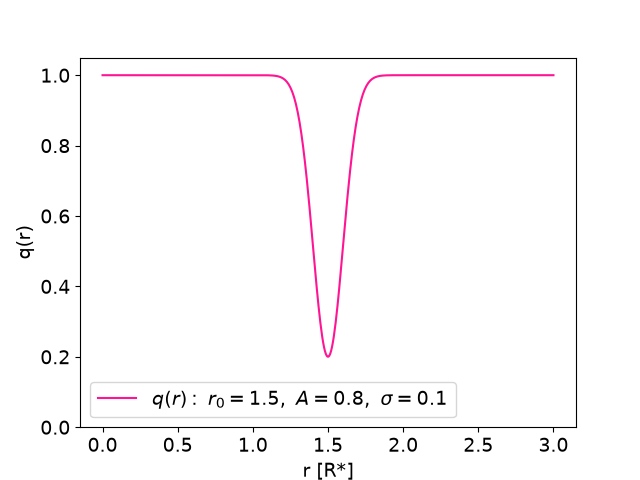}
    \caption{Shape of the added heating potential (Trial 1,
    Eq.~\eqref{eq:q(r)_0})
    with the parameters given in the plot.
    $\frac{I_\text{H}}{m_\text{H}}$ is taken as 1 here for plotting.}
    \label{Fig:q_r_1_shape}
\end{figure}

Trial 1: For the initial trial with the additional heating term in the specific internal energy $e(r)$, we test a ``1 minus Gaussian dip" function of the radius $r$, given by 
\begin{equation} \label{eq:q(r)_0}
    q(r) = \frac{I_\text{H}}{m_\text{H}} \left[ 1 - \mathrm{A} \exp{\left(-\frac{(r-r_0)^2}{2\sigma^2} \right)} \right],
\end{equation}
which gives a similar distribution to the expected hydrogen ionization fraction (the shape
is
plotted in Fig. \ref{Fig:q_r_1_shape}). 
The shape of this function is motivated by a recombination zone found in the A supergiant models by \citet{acmfkont},
and the effect of this form of added heating on the outflow is discussed in Section \ref{sec:results}.
The parameters A, $r_0$, and $\sigma$ can be varied to adjust the depth, radial location and width of the inverted curve, respectively.

\begin{figure}
[t]
    \centering
    \includegraphics[width=0.8\linewidth]{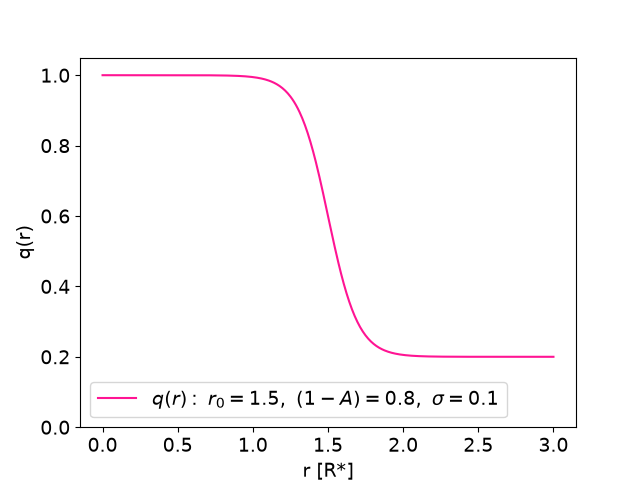}
    \caption{Shape of the added heating potential (Trial 2,
    Eq.~\eqref{eq:q(r)_1}) with the parameters given in the plot.
    $\frac{I_\text{H}}{m_\text{H}}$ is taken as 1 here for plotting.}
    \label{Fig:q_r_sigmd_shape}
\end{figure}

Trial 2: For the second trial, we use a sigmoid-like heating function given by
\begin{equation}\label{eq:q(r)_1}
q(r) = \frac{I_\text{H}}{m_{H}} \left( A + \frac{1-A}{1+\exp{(\frac{r-r_0}{\sigma})}} \right),
\end{equation}
where $(1-A)$ is the depth of the profile, $r_0$ is the location, $\sigma$ is the width (the shape is shown in Fig. \ref{Fig:q_r_sigmd_shape}). 
The shape of this function corresponds to recombination that appears in the photosphere.

We also note its derivative here,
\begin{equation}\label{eq:dqdr_sigmoid}
    \frac{dq}{dr} = q'(r) = -
    \frac{I_\text{H}}{m_{H}}
    (1-A) \frac{\exp(\frac{r-r_0}{\sigma})}{\sigma\left[ 1 + \exp{(\frac{r-r_0}{\sigma})}\right]^2}, 
\end{equation}
for later reference.

\subsection{Finding a solution}\label{subsec:solution_search}

Now, 
Eq.~\eqref{eq:vel_1a}
can be rewritten using 
Eq.~\eqref{eq:dcs2/dr}
as,
\begin{equation}
    \left( v -\frac{c_s^2}{\gamma v} \right) \frac{dv}{dr} = \frac{2c_s^2}{\gamma r} + \frac{(\gamma -1)}{\gamma} \left( \frac{dq}{dr} + v \frac{dv}{dr} + \frac{GM}{r^2}\right) - \frac{GM}{r^2},
\end{equation}
or 
\begin{equation}\label{eq:vel_final}
    \frac{(v^2 - c_s^2)}{v} \frac{dv}{dr} = \frac{2c_s^2}{r} + (\gamma -1) \frac{dq}{dr} - \frac{GM}{r^2},
\end{equation}
which gives us a single first order ODE for velocity, in terms of $r$, $q(r)$ and $c_s(r,v)$ (see Eq. \ref{eq:c_s^2__1}). We are looking for a transonic solution where the velocity passes from a subsonic to a supersonic value at a critical radius $r_\mathrm{crit}$ from centre of the star. So the solution should satisfy the following condition at this particular radius, 
\begin{equation}\label{eqn:vel_at_rc}
   v = c_s \implies \frac{2c_s^2}{r} + (\gamma - 1) \frac{dq}{dr} - \frac{GM}{r^2} = 0,
\end{equation}
or
\begin{equation}\label{eq:crit_rad1}
     \frac{2c_\mathrm{crit}^2}{r_\mathrm{crit}} - \frac{GM}{r_\mathrm{crit}^2} + (\gamma - 1) q'(r_\mathrm{crit}) = 0.
\end{equation}

In order to find a transonic solution for the velocity equation, we need to 
choose a physically valid adiabatic index $\gamma$ for our model stellar atmosphere. The parameter space for this was explored in a recent work by \citet{Westrichetal_2026MNRAS.548ag755W}.
Furthermore, we need to compute the value of the Bernoulli constant $B$ (from Eq. \ref{eq:energy2}) at the base of the wind by assuming the relevant stellar parameters. Here, we can turn to the HR diagram we plotted in Section \ref{sec:hrdiag-recombination}. From our estimate, a subset of both the extended red giant and AGB
stars has enough energy from recombination for their atmospheric hydrogen to overcome the gravitational potential energy. So we take a single chosen AGB star's evolutionary point parameters from this subset of MIST data to look for a solution.

For both our analytical and hydrodynamical models, we take a star with
mass $M_* = 1.052 ~\mathrm{M}_\odot$, radius $R_* = 280.2~\mathrm{R_\odot}$, and temperature $T_\text{eff} = 2862~\mathrm{K}$.
We assumed the base density of the wind to be $\rho = 10^{-14}~\text{g cm}^{-3}$. The subsonic base wind velocity is taken as a small fraction of the local sound speed. We tested the model with different values of the adiabatic index $\gamma$ including 1.0001, 1.02 (both nearly isothermal) and 1.2. 

The value of Bernoulli constant is estimated by computing the Bernoulli integral (Eq. \ref{eq:energy2}) at the wind base with the above values. The critical radius ($r_\text{crit}$) can then be estimated by any suitable root-finding method, like Newton-Raphson, with Eq.~\eqref{eq:crit_rad1}.
The algorithm for finding a solution for the velocity equation (Eq. \ref{eq:vel_final}) is detailed in Appendix \ref{sec_ap:root_finding_algo}. Further details of the solution we obtained are described in Section \ref{subsec:anal_results}.


\subsection{
Time-dependent
modelling with CASTRO code}\label{subsec:CASTRO_modeling}

To further test the validity and numerical stability of the steady-state solution, we simulated the system in the time-dependent hydrodynamical code CASTRO \citep{CASTRO_I_2010ApJ...715.1221A, CASTRO_0_JOSS_Almgren2020}. The code solves the full compressible Euler equations in conservative form on an Eulerian grid. The code can optionally include radiation as well, which will be useful in our future implementations. Our intention here is to test our current model for hydrodynamical convergence and stability.

For our 1D spherical coordinates, with added source terms for gravity and 
heating,
CASTRO essentially evolves the system of coupled partial differential equations;
the equations of continuity
\begin{equation}\label{eq:castro_mass_cons}
    \frac{\partial\rho}{\partial t} = 
    -\frac{1}{r^2} \frac{\partial}{\partial r} (r^2 \rho v),
\end{equation}
momentum conservation,
\begin{equation}\label{eq:castro_mom_cons}
    \frac{\partial }{\partial t} (\rho v) = 
    -\frac{1}{r^2} \frac{\partial }{\partial r} (r^2 \rho v^2) 
    - \frac{\partial P}{\partial r} + \rho g(r) + 
    \mathbf{S_{\mathrm{ext,} \rho v}} ,
\end{equation}
and total energy conservation,
\begin{equation}\label{eq:castro_energy_cons}
    \frac{\partial}{\partial t} (\rho E) = 
    -\frac{1}{r^2} \frac{\partial}{\partial r} [r^2 (\rho E + P) v ] 
    + \rho v g(r) + \mathbf{S_{\mathrm{ext}, \rho E}} ,
\end{equation}
where, $E = e + \frac{v^2}{2}$ is the total specific energy, and $g(r) = -\frac{GM}{r^2}$. 
The heating source term $q(r)$ enters the energy equation as 
\begin{equation}\label{eq:heat_term_castro_energy_source}
    \mathbf{S_{\mathrm{ext}, \rho E}} = -\rho v
    \frac{dq}{dr}.
\end{equation}

CASTRO evolves the conserved density values $\textbf{U} (\rho, \rho v, \rho E)$ and we can extract the coordinate dependent density ($\rho$), velocity ($v$), and specific internal energy ($e$) profiles at any given time from these. 
We can then estimate the temperature, pressure, sound speed and other relevant quantities using the chosen equation of state. 
With the gamma-law equation of state, we can estimate, $T = \frac{\mu m_\text{H}}{k_B} (\gamma -1) e$, $P = (\gamma - 1) \rho e$, and $c_s = \sqrt{\frac{\gamma P}{\rho}} = \sqrt{ \gamma (\gamma -1)e}$.

First of all, in order to validate that the code works properly for our purpose, we started out testing the classical Parker model in CASTRO. We kept the default $\gamma$-law equation of state with $\gamma = 1.0001$ to approximate for the isothermal case. We used a resolution of 4096 cells in the $r$ direction without adaptive mesh refinement. The numerical domain ranged from $r = R_0= 1 R_*$ to $r = R_\text{out} = 100 R_*$. We initialized this model with a constant temperature over the entire domain ($r = R_0$ to $R_\text{out}$) which is the base wind temperature. The initial velocity $v_0$ was also kept a constant across the whole domain and it was chosen to be a small subsonic fraction of the base isothermal sound speed $a_0$, $v_0 = 10^{-4} a_0$. The initial density over the domain was set to be decreasing with radius, $\rho (r) \propto 1/r^2$. For testing this isothermal Parker model, we have used the following stellar parameters: $M_* = 1~\text{M}_\odot$, $R_* = 1 ~\text{R}_\odot$, $T_\text{base} = 1.5 \times 10^6~\text{K}$, and $\rho_\text{base} = 10^{-14} \text{g cm}^{-3}$. Gravity is implemented as an extra source term to the momentum and energy equations (as given in Eqs. \ref{eq:castro_mom_cons} and \ref{eq:castro_energy_cons}). It is implemented this way because the built-in point-mass gravity module in CASTRO requires the numerical domain to start at the origin, $r= 0$, which is not our case.  We ran this model for a long enough time (up to $\sim 10^9 \text{s}$ in the star's frame of reference) and noted that it reached the analytical Parker velocity profile for the given parameters and continued in that stationary profile after that.

To test our model with the added heating term, we built upon the above initial condition. For this model too, we initialized the simulation with a small flat subsonic velocity profile over the entire domain ranging from $1~\text{R}_*$ to $100~\text{R}_*$. This initial value was set to be $2.32 \times10^{-4} a_0$.
We let the code evolve the system to the expected final solution over time from the given initial conditions, as it is done in time-dependent codes. We used the same stellar parameters we used for our analytical solution to initialize this model in CASTRO. The values are $M_* = 1.052 ~\mathrm{M}_\odot$, $R_* = 280.2~\mathrm{R_\odot}$, and $T_\text{eff} = 2862~\mathrm{K}$. The temperature was initialized at this given constant value over the entire grid. The density at the base of the wind, which is at $r= 1~\text{R}_*$, was set to be $\rho_\text{base}=10^{-14} \text{g cm}^{-3}$. The added heating term (Eq. \ref{eq:q(r)_1}) enters the system through the energy equation as given in Eq.~\eqref{eq:heat_term_castro_energy_source}.
Since CASTRO uses a dual-energy scheme by simultaneously and separately evolving the internal energy ($\rho e$) and total energy ($\rho E$), we had to source both of these with the added heating term. The code uses its own algorithm to properly evolve the current internal energy from these variables to avoid numerical inconsistencies during extreme Mach number scenarios. 
We note here that the derivative given in Eq.~\eqref{eq:dqdr_sigmoid} is always negative. This means, for the right hand side of the Eq.~\eqref{eq:heat_term_castro_energy_source} to be positive, the velocity always needs to be positive or outward directed from the centre of the star. If the velocity becomes negative at any point in time, the added heating source term would essentially become a cooling term in our model. To ensure that the velocity stays positive during the simulation, we needed to make sure that we initialize our model in hydrostatic equilibrium. So we prescribe an initial density profile of the 
form $\rho(r) = \rho_\text{base} \exp{\left[\frac{GM}{a_0^2} \left( \frac{1}{r} - \frac{1}{R_*} \right)\right]}$. 
This ensures that a non-negative velocity is maintained near the star during the initial time steps.

Since the heating term we introduce in Eq.~\eqref{eq:q(r)_1} is a highly localized potential in $r$, we found that it can cause numerical overflows in CASTRO during the time evolution. To avoid this issue, we used a time ramp function for gradually introducing the heating term in the initial stage of the simulation. Instead of directly introducing $\mathbf{S}_{\text{ext}, \rho E}  = Q_\text{vol} \equiv -\rho v_r \frac{dq}{dr}$ at $t=0$, we define a time ramp function $f(t)$ to introduce the heating term gradually so that $\mathbf{S}_{\text{ext}, \rho E}  = f(t) Q_\text{vol}$. The time ramp function is defined as $f(t) = \frac{1}{2} [1 - \cos(\pi t/t_\text{ramp})]$ for $0 \leq t < t_\text{ramp}$ and $f(t) = 1$ for $t\geq t_\text{ramp}$, where $t_\text{ramp} = n_\text{ramp} \times \frac{R_*}{a_0}$ with $n_\text{ramp} = 20$. This made sure that the simulation doesn't crash due to numerical errors resulting from the highly localized initial heating term. The results from these models are described in 
Section \ref{subsec:CASTRO_results}.

\section{Results and discussion}\label{sec:results}

\subsection{Wind solution from the 
stationary
model}\label{subsec:anal_results}

\begin{table*}
[t]
    \centering
        \caption{
        Parameters of the resulting solution for heating function from Eq.~\eqref{eq:q(r)_1}.
        The solutions in the table were calculated for a star with the parameters: mass $M_* = 1.052 ~\mathrm{M}_\odot$, radius $R_* = 280.2~\mathrm{R_\odot}$, and temperature $T_\text{eff} = 2862~\mathrm{K}$.}
    \begin{tabular}{ccccccc}\toprule
         &  &  &  \multicolumn{2}{c}{$\gamma = 1.02$}&  \multicolumn{2}{c}{$\gamma = 1.2$}\\
         \cmidrule(lr){4-5} \cmidrule(lr){6-7}
         $r_0$ [$\text{R}_*]$&  $1-A$&  $\sigma ~[\text{R}_*]$&  $r_\text{crit} ~[\text{R}_*]$&  $c_\text{crit}$ [km/s]&  $r_\text{crit} ~[\text{R}_*]$& $c_\text{crit}$ [km/s]\\\midrule
         1.5&  0.4&  0.15&  1.785&  30.03&  1.767& 88.10\\
         &  &  0.20&  1.835&  28.81&  1.807& 83.99\\
         &  &  0.25&  1.882&  27.64&  1.841& 79.99\\
         &  0.6&  0.15&  1.771&  36.27&  1.765& 107.81\\
         &  &  0.20&  1.815&  34.69&  1.804& 102.74\\
         &  &  0.25&  1.855&  33.18&  1.838& 97.80\\
         &  0.8&  0.15&  1.765&  41.58&  1.765& 124.44\\
         &  &  0.20&  1.805&  39.70&  1.803& 118.56\\
 & & 0.25& 1.841& 37.90& 1.836&112.84\\
 2.0& 0.4& 0.15& 2.310& 30.86& 2.305&91.25\\
 & & 0.20& 2.354& 30.21& 2.346&89.14\\
 & 0.6& 0.15& 2.302& 37.42& 2.304&111.75\\
 & & 0.20& 2.342& 36.57& 2.344&109.15\\
 & 0.8& 0.15& 2.298& 42.99& 2.303&129.04\\
 & & 0.20& 2.336& 41.97& 2.343&126.02\\
 2.5& 0.4& 0.15& 2.838& 31.15& 2.839&92.36\\
 & & 0.20& 2.891& 30.66& 2.892&90.83\\
 & 0.6& 0.15& 2.833& 37.83& 2.838&113.15\\
 & & 0.20& 2.883& 37.20& 2.891&111.27\\
 & 0.8& 0.15& 2.830& 43.50& 2.838&130.67\\
 & & 0.20& 2.879& 42.75& 2.890&128.49\\ \bottomrule 
    \end{tabular}
    \label{tab:param_scan_q(r)_sigmoid}
\end{table*}

We checked for possible solutions of the velocity equation (Eq. \ref{eq:vel_final}) with two different added heating terms, as described in Section \ref{subsec:heating_func}. 
We obtained the radial velocity profile in each trial with varying parameters of the heating function, following the algorithm described in Appendix \ref{sec_ap:root_finding_algo}.

\begin{figure}
[t]
    \centering
    \includegraphics[width=1.0\linewidth]
    {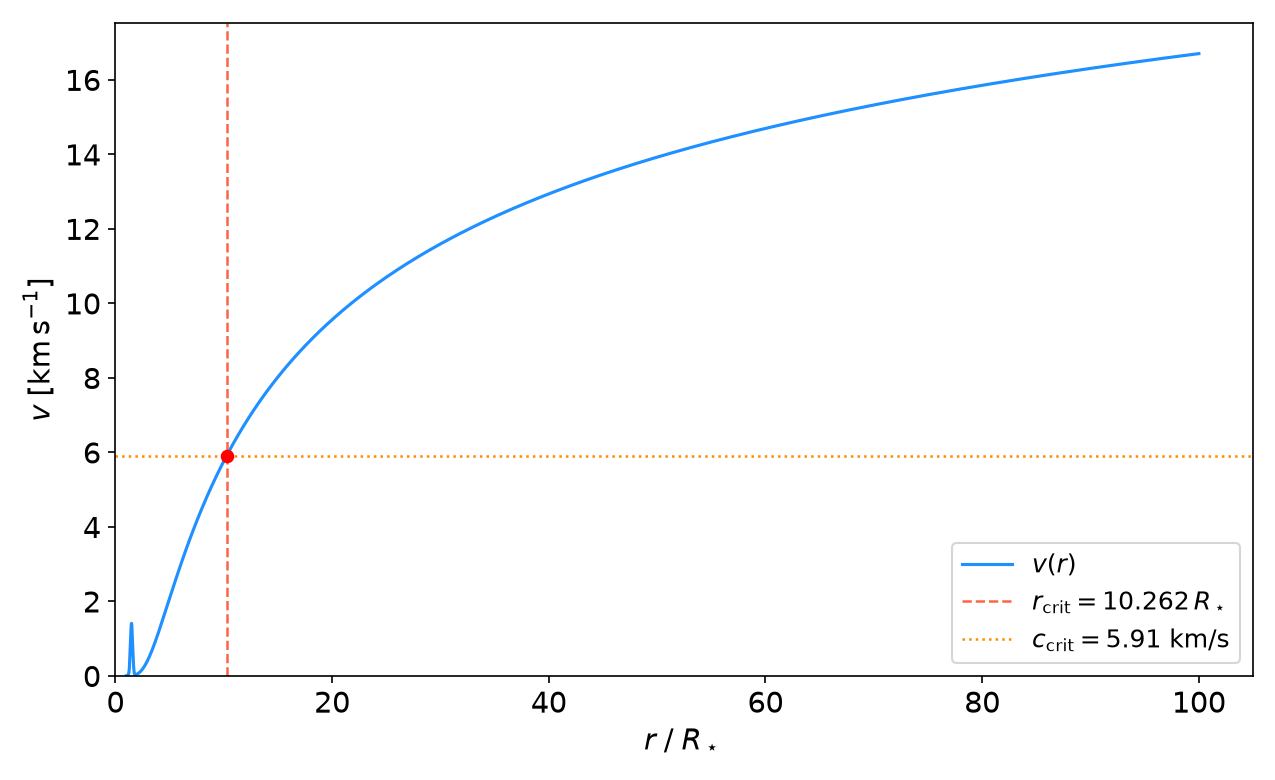}
    \caption{
    Radial velocity profile for the modified velocity equation with the heating function from Eq.~\eqref{eq:q(r)_0} with the parameters $r_0 = 1.50\,R_*$, $A = 0.80$, and $\sigma = 0.10\,R_*$ and the stellar parameters $M=1.052\,M_\odot$,  $R=280.2\,R_\odot$,  $T_0=2862\,\text{K}$. We used $\gamma = 1.02$ for this model. The plot shows the classical Parker solution with an additional spike in velocity at the location of the added heating term.
    }
    \label{fig:example_soln_1_q(r)_0}
\end{figure}

The initial trials were carried out with a heating term
following the hydrogen ionization fraction in A-supergiants \citep{acmfkont} as given in 
Eq.~\eqref{eq:q(r)_0}. The resulting velocity profile for this type of heating term is shown in 
Fig.~\ref{fig:example_soln_1_q(r)_0}
along with the parameters used for the heating term. The plot essentially shows the velocity profile of the classical Parker solution for the given stellar parameters, with an additional velocity spike at the location of the added heating. With this, we confirmed that the added heating term can modify the solution, and adjusting its parameters can alter
the sonic point.

\begin{figure}
    \centering
    \includegraphics[width=1.0\linewidth]{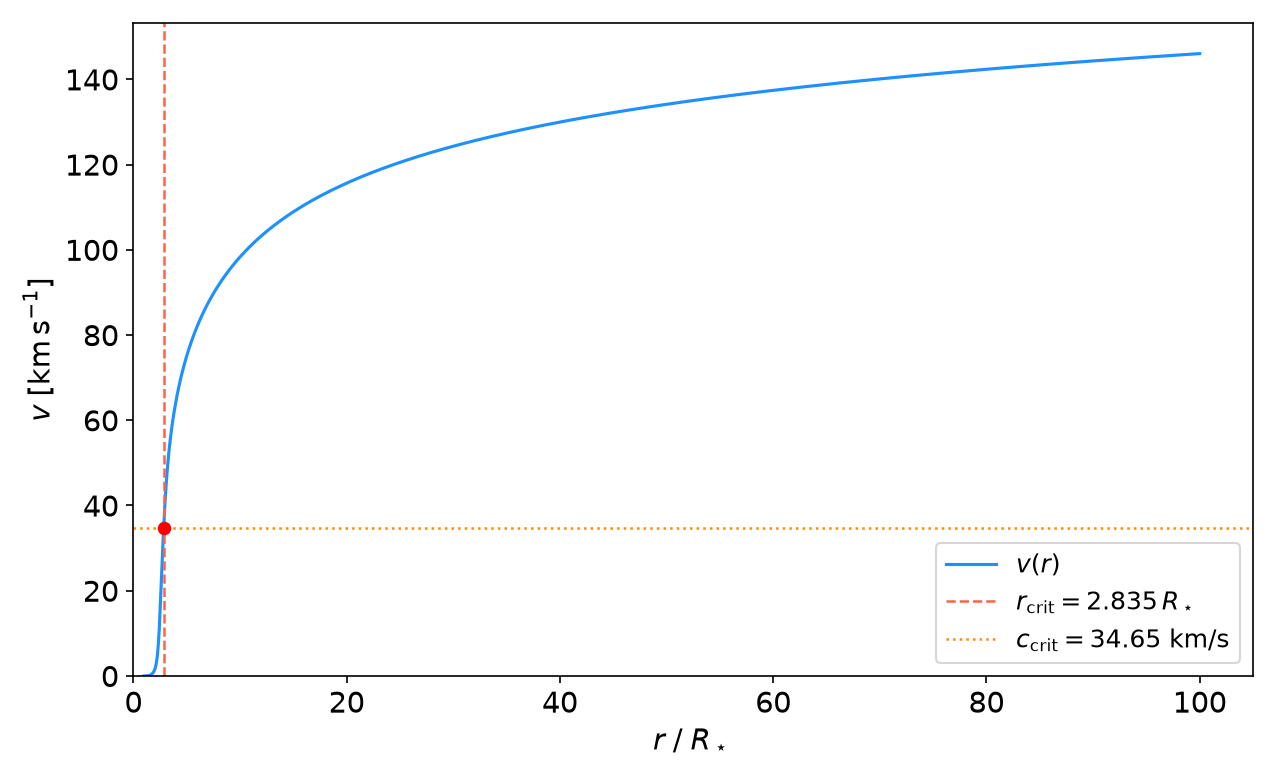}
    \caption{
    Radial velocity profile of the wind with heating function from Eq.~\eqref{eq:q(r)_1}
    with the parameters $r_0 = 2.50\,R_*$, $(1-A)=0.50$, and $\sigma = 0.15\,R_*$. 
    The stellar parameters remain the same as in Fig. \ref{fig:example_soln_1_q(r)_0}. 
    We used $\gamma = 1.02$ for this specific model.}
    \label{fig:example_soln_1_q(r)_1}
\end{figure}

We further tested the heating term from Eq.~\eqref{eq:q(r)_1}, which is motivated by the heating resulting from hydrogen recombination in a star's atmosphere. The resulting wind solution for this physically motivated sigmoid-like heating profile is given in Fig.~\ref{fig:example_soln_1_q(r)_1}. The parameters used for this particular solution are given in the plot. This solution clearly has its critical point much closer to the star than the classical Parker solution. The critical velocity and the maximum velocity values attained by the wind far from the star are also an order of magnitude higher than the classical isothermal Parker wind. A parameter scan of the heating function for this class of solutions and the resulting velocity values is given in Table \ref{tab:param_scan_q(r)_sigmoid}. 
From the data, we can immediately notice the following characteristics of the solution. Moving the location of the extra heating ($r_0$) closer to the centre of the star brings the critical radius ($r_\text{crit}$) also nearer to the star's centre. Increasing the amplitude ($1-A$) of the heating function also brings the sonic point closer to the centre of the star. Both of these conditions, in turn, increase the critical velocity. On the other hand, increasing the width ($\sigma$) of the function moves the sonic point farther from the star's centre. It is also worth noticing that using a higher adiabatic index ($\gamma = 1.2$ compared to the near-isothermal value $\gamma = 1.02$) only slightly affects 
location of 
the critical point while increasing the critical velocity significantly. 
These variations follow from the properties of the Parker wind solution \citep{Lamers_1999isw..book.....L}.
The 
mass-loss rate
of recombination-driven winds
is discussed in Section \ref{sec:mass-loss-estimate}. 

\subsection{CASTRO results}\label{subsec:CASTRO_results}

\begin{figure*}
    \centering
    \includegraphics[scale=0.4]{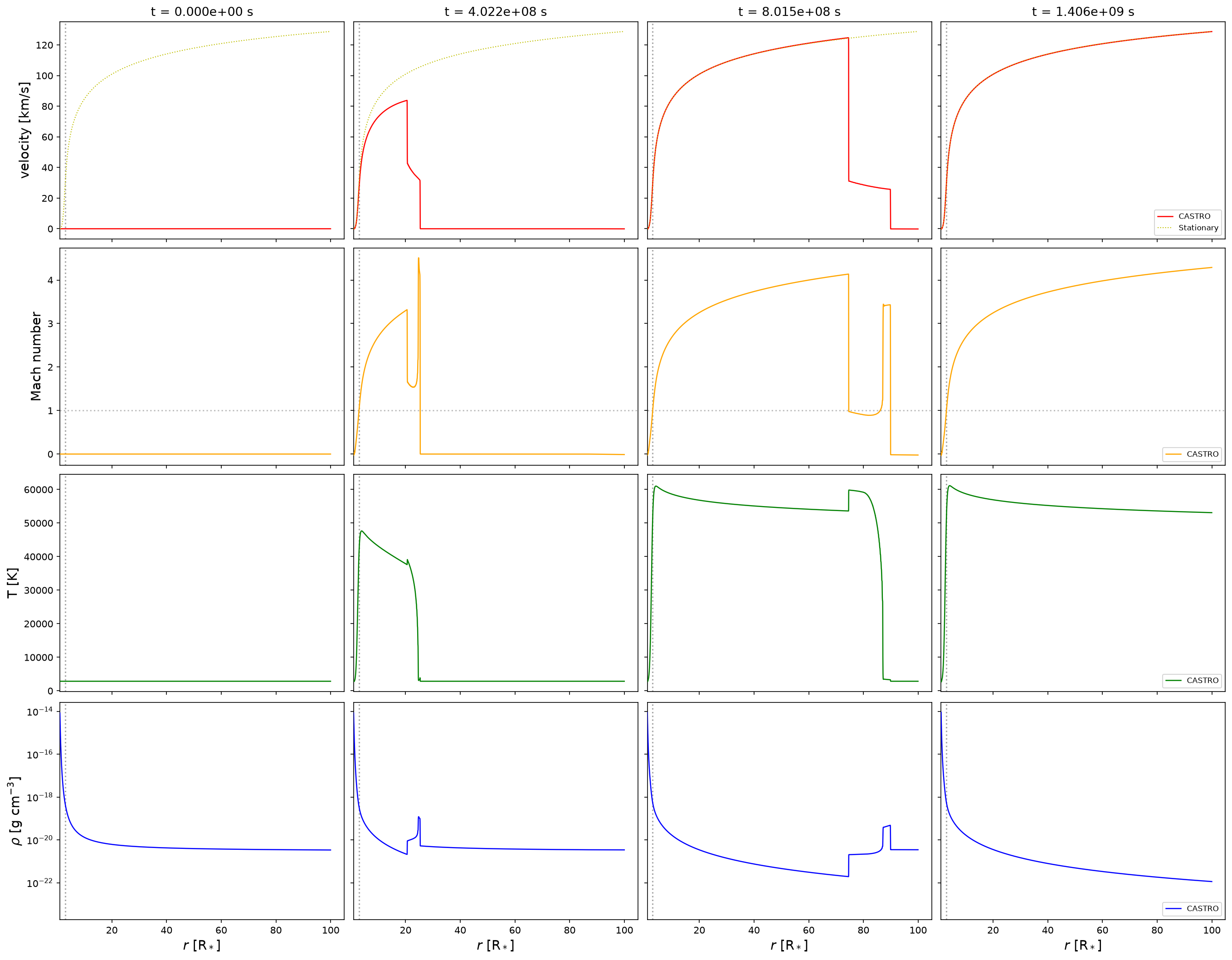}
    \caption{
    Radial profiles of velocity, Mach number, temperature, and density at four chosen time-steps
    (denoted in the graph) from a selected model of CASTRO simulations. This model was initialised with the added heating term given in Eq.~\eqref{eq:q(r)_1} and the initial profiles shown in the leftmost column. The stellar parameters remain the same as in Fig. \ref{fig:example_soln_1_q(r)_0}. The parameters of the heating term are $r_0 = 2.50\,R_*$, $(1-A) = 0.40$, and $\sigma = 0.30\,R_*$. We used $\gamma = 1.02$ for this model.
    The rightmost column shows the evolved stationary state values, which 
    nicely agree with the results of the stationary model plotted in the upper panel.
    The vertical dotted line in all the columns marks the sonic point for this particular model.
    }
    \label{fig:castro_results}
\end{figure*}

Figure \ref{fig:castro_results} shows results from time-dependent simulations obtained using the
CASTRO code for a chosen model with the added heating term in Eq.~\eqref{eq:q(r)_1}. This model used the same stellar parameters as 
our 
stationary
model in Sect.~\ref{subsec:anal_results}.
The added heating was located at $r_0 = 2.5 ~\text{R}_*$, with an amplitude of $(1-A) = 0.4$ and a width of $\sigma = 0.30~\text{R}_*$. 
The first column from the left of the figure shows the initial profiles as described, at time $t = 0~\text{s}$, with zero initial velocity and Mach number values across the domain. The initial temperature in the domain shows a constant value of 2862 K. The initial density follows the distribution described in Section \ref{subsec:CASTRO_modeling} maintaining the initial hydrodynamic equilibrium. 
The second and third columns in the figure show the evolution towards the final solution. The plots show the velocity growing from zero to the predicted stationary solution values at a speed given by the sound crossing time in the domain. The Mach number also follows this pattern and mechanical shocks are also apparent from the plots at the expanding front. The temperature also gradually rises across the entire domain to about 60\,000\,K drawing energy from the added heating. The velocity gradually settles to a radial dependent lower value following the mass conservation. 
Both the temperature and density distributions also suggest the expected shock at the expanding wind front. 
The rightmost column in Fig.~\ref{fig:castro_results} shows the final stationary evolved state.
The values match that predicted by the analytical solution and they stay stationary as long as we ran the simulation after this. This confirmed the numerical convergence and stability of the solution. 

The above model ran in CASTRO confirmed the hydrodynamic consistency of our model for recombination driven stellar winds. In the future, we plan to investigate our model further by implementing the radiation module in CASTRO for solving the radiation hydrodynamic equations for this problem, which would further examine the physics described by our initial model.

\subsection{Comparison with other wind driving models}

The recent work by \citet{Yang_2026arXiv260619422Y} addresses a similar question to ours while using a different methodology. They mainly suggest that recombination can enhance or unbind a flow that is already expanding but rarely supports a fresh wind launching from hydrodynamic equilibrium. We further discuss the conditions for plausible recombination-driven winds in Section \ref{sec:mass-loss-estimate}.  
Direct wind simulations that followed the acceleration region above the stellar surface suggested instabilities and clumped outflows \citep{Gomez_Williams_2003MNRAS.344..725G, SudquistOwockiPuls_2018A&A...611A..17S}. This encourages further time-dependent hydrodynamical simulations where our model can serve as a starting point for the recombination-driven winds. 
While models already exist for pulsation driven, recombination assisted wind models in evolved stars \citep{WachterSchroder_2002A&A...384..452W}, our model investigates recombination as the main wind driver. 
There are also three-dimensional hydrodynamical common envelope simulations of a red-supergiant donor, run with different recombination energy treatments, which suggest an impulsive asymmetric common envelope ejection \citep{Lau_etal_2022MNRAS.516.4669L}. In their adiabatic models, helium recombination increased the final separation up to 16\% while hydrogen recombination increased the unbinding further. This also motivates further investigations for recombination-driven wind models.

\section{Mass-loss rate estimate}\label{sec:mass-loss-estimate}

Hydrodynamical equations describing the recombination-driven winds 
do not explicitly depend on density. Therefore, the
hydrodynamical equations elaborated here
can in principle describe outflows with arbitrary strength. This means 
that we have to add additional constraints that would determine 
the wind mass-loss rate.

Recombination-driven winds are possible thanks to addition of the
heating potential $q(r)$ to hydrodynamical equations. Therefore, 
this term is a key to estimate the wind mass-loss rate. The
heating term describes the energy released as a result of 
recombination, which is given by the ionization potential and a
fraction of atoms that recombined. These parameters are constant
for a complete recombination; therefore, it is the radial
dependence of $q(r)$ that determines the mass-loss rate.

Radial variations of the heating function are given by
recombination. The location where the recombination takes place
depends on the electron density. This could be the desired effect 
that determines the mass-loss rate. With increasing wind density
the location of recombination region moves towards the star. This
would be important when the location of this region moves too
deep into the stellar envelope, into the regions where the
recombination itself is not able to drive the wind. However, the
location of the recombination region depends more strongly on the 
temperature than on the electron density, and consequently, the
change of $q(r)$ with density may not be strong enough to 
determine the mass-loss rate.

This could mean that the mass-loss rate is not constrained by the
wind equations themselves but by the speed at which the star is
able to replenish the material lost, that is, by the evolutionary
time scale. In this sense, the recombination-driven winds could be
analogous to outflowing disks of critically rotating stars, where
the mass-loss rate is given by the requirement to keep the star at
(or slightly below) the critical rotation rate \citep{kom}. In
principle, the mass-loss rate due to the recombination-driven wind
can be implemented in evolutionary calculations by peeling all
the mass whose gravitational potential energy per particle is
lower than the ionization energy, in analogy to the mass-loss by 
outflowing disks \citep{Granada_2013A&A...553A..25G}.

Variation of the fraction of the released recombination energy
that is deposited to heat can provide another effect that modulates
the mass-loss rate. Recombination can proceed either radiatively
or collisionally. Within radiative recombination, the thermal
energy of the recombining electron is transformed into radiation,
taking the heat away. In the case of collisional recombination, 
the recombination energy is carried out by the free electron, 
releasing heat. The ratio of energy released by collisional and
radiative recombination is \citep{kpp}
\begin{equation}
\varepsilon=\frac{Q_\text{c}^\text{H}}{Q_\text{bf}^\text{C}},
\end{equation}
where the energy rate released by collisional recombination is
\begin{equation}
Q_\text{c}^\text{H}=n_\text{e}\left(\frac{n_1}{n_2}\right)^*
n_2q_{12}(T)\,h\nu_{12},
\end{equation}
and the energy rate taken away by radiative recombination is
\begin{equation}
Q_\text{bf}^\text{C}=4\pi\left(\frac{n_1}{n_2}\right)^*
n_2\int_0^\infty\alpha_\text{bf,12}(\nu)
\left(J_\nu+\frac{2h\nu^3}{c^2}\right)e^{-h\nu/kT}
\left(1-\frac{\nu_{12}}{\nu}\right)\,\text{d}\nu.
\end{equation}
We accounted for just the ground level with number density
$n_1$ and assumed unity occupation probabilities. Here $n_\text{e}$
is the number density of free electrons, 
$(n_1/n_2)^*$ denotes population ratio of neutral and ionized atoms
in local thermodynamic equilibrium, $n_2$ is the number density of
ionized atoms, $q_{12}(T)$ is the collision strength, $\nu_{12}$
is the frequency of ionization edge, $\alpha_\text{bf,12}(\nu)$
is photoionization cross-section, and $J_\nu$ is the mean
intensity.

Neglecting the stimulated recombination ($J_\nu=0$), approximating
the photoionization cross-section by Kramers formula 
\citep{hubenymihalas} $\alpha_\text{bf,12}(\nu)\approx\alpha_{12}/
v^3$ with $\alpha_{12}$ being constant related to the edge cross-section
of $\alpha_{12}/\nu_{12}^3$, and replacing
the photoionization integral by its first-order estimate we derive
\begin{equation}
\varepsilon=\frac{n_\text{e}q_{12}(T)h\nu_{12}c^2}
{8\pi\alpha_{12}kT}e^{\frac{h\nu_{12}}{kT}}.
\end{equation}
Approximating the collision strength by the Seaton's formula
\citep{hubenymihalas} 
\begin{equation}
q_{12}(T)=1.55\times10^{12}T^{-1/2}
\frac{\alpha_{12}}{\nu_{12}^3}e^{-\frac{h\nu_{12}}{kT}}\frac{kT}{h\nu_{12}}
\end{equation}
we finally arrive at 
\begin{equation}
\varepsilon=6.17\times10^{10}\frac{n_\text{e}c^2}
{T^{1/2}\nu_{12}^3}=5\times10^{-7}
\left(\frac{n_\text{e}}{10^{10}\,\text{cm}^{-3}}\right)
\left(\frac{T}{1000\,\text{K}}\right)^{-1/2}.
\end{equation}
This indicates that relatively high densities are required to
release the recombination energy as heat (corresponding to 
$\varepsilon\gtrsim1$). Assuming that 
recombination appears close to the sonic point of a giant 
with a
radius of the order of a hundred solar radii, recombination-driven
winds require mass-loss rates of the order of 
$0.1\,M_\odot\,\text{yr}^{-1}$ or higher.

\section{Summary and conclusions}\label{sec:conclusions}

In this work, we devised an initial model for stellar winds driven by hydrogen recombination, based on the classical Parker solution for the solar wind. We started out by checking the possibility of recombination-driven winds across the HR diagram. 
By comparing the gravitational potential energy available per hydrogen atom to the hydrogen ionisation energy in stars across the HR diagram, we showed that stellar winds driven by recombination are possible in red giants and AGB stars.

Then we developed a basic analytical model with an added heating potential to the classical Parker solution and moved on to checking particular solutions for the resulting velocity equation.
The added heating term, analogous to the heat released by recombination in the atmospheres of certain stars, can drive an outflow from a star with velocities up to an order of magnitude higher than the classical Parker wind.

We implemented the model in the time-dependent
astrophysical simulation code CASTRO and tested the solution further by evolving the model in time.
Our CASTRO simulations demonstrate that recombination driven outflows can possibly be launched from hydrostatic equilibrium, given the right conditions. This can be further tested by future radiation hydrodynamic simulations for which our model serves as a starting point. The simulations also verify the stationary solution we obtained, and its stability as long as the heating term remained the same. 

The mass-loss due to recombination driven winds strongly depends on the radial distribution of the heating function, which is given by recombination itself. The mass-loss rate is not constrained by the wind equations themselves but instead by the rate at which a star is able to replenish the material lost, which in turn depends on its evolutionary stage.  
We emphasized the principal difference between the radiative and collisional recombination. While the radiative recombination deposits the ionization energy to the radiation field and locally contributes to cooling, the collisional recombination doposits the energy locally as a heat. This allowed us to estimate
the minimum recombination-driven wind mass-loss rate by requiring that most of the recombination energy is deposited in the atomic thermal energy. This condition gives the minimum mass-loss rate of the order of $
0.1
\,M_\odot\,\text{yr}^{-1}$.

\begin{acknowledgements}
This work was supported by the grant GA \v{C}R 25-15910S. We thank 
Drs.~Ji\v r\'\i\ Kub\'at and Ond\v rej Pejcha for the discussion
of radiative heating and the role of recombination in envelope
ejection. 
We thank Dr. Petr \mbox{Kurfürst} for providing a nice introduction to the CASTRO code.
The computational part of this work was supported by the Ministry of Education, Youth and Sports of the Czech Republic through e-INFRA CZ (ID:90254).
\end{acknowledgements}

\appendix
\nolinenumbers

\section{Algorithm for solving the modified Parker velocity equation}
\label{sec_ap:root_finding_algo}

We can rewrite Eq.~\eqref{eq:vel_final} as
\begin{equation} \label{eq:vel_ode_apend}
    \frac{dv}{dr} = \frac{v}{v^2 - c_s^2(r,v)} \left[ \frac{2c_s^2(r,v)}{r} - \frac{GM}{r^2} + (\gamma - 1) q'(r) \right],
\end{equation}
which gives a single first order 
ordinary differential equation
in $v(r)$ only.
We used the following algorithm
to arrive numerically at a solution of this equation:\newline

\noindent \textbf{Step 1:} Calculate the Bernoulli constant B:\newline

Evaluate the Bernoulli integral 
Eq.~\eqref{eq:bernoulli}
at the base of the wind with the base values, radius $r = R_*$, $T = T_0$ the base temperature, $\rho = \rho_0$ the base density, and the base velocity $v = v_0 \ll a_0$, where 
$a_0^2 \equiv c_s(r=R_*) = \frac{\gamma k_B T_0}{\mu m_\text{H}}$.
This gives
\begin{equation}
    B = \frac{v_0^2}{2} + \frac{a_0^2}{\gamma - 1} + q (r_0) - \frac{GM}{r_0} \approx \frac{a_0}{\gamma - 1} + q(r_0) - \frac{GM}{r_0}
\end{equation}

\noindent \textbf{Step 2:} Use a root finding (Newton-Raphson or similar) method to find $r_\mathrm{crit}$:\newline

At the critical radius $r_\mathrm{crit}$, two conditions must be met,

\textbf{I}:
When $v = c_\mathrm{crit}$, the LHS of Eq.~\eqref{eq:vel_final} vanishes,
\begin{equation}\label{eq:conditionA}
    c_\mathrm{crit}^2 = \frac{GM}{2r_\mathrm{crit}} - \frac{(\gamma-1)}{2} r_\mathrm{crit} q'(r_\mathrm{crit}).
\end{equation}

\textbf{II}: The Bernoulli equation 
\eqref{eq:bernoulli}
evaluated for $v=c_\mathrm{crit}$ is
\begin{equation}
    \frac{1}{2} c_\mathrm{crit}^2 + \frac{c_\mathrm{crit}^2}{\gamma -1} + q(r_\mathrm{crit}) - \frac{GM}{r_\mathrm{crit}} = B,
\end{equation}
which gives,
\begin{equation}\label{eq:conditionB}
     c_\mathrm{crit}^2 = \frac{2(\gamma -1)}{\gamma + 1} \left[ B - q(r_\mathrm{crit}) + \frac{GM}{r_\mathrm{crit}} \right].
\end{equation}

Equating Eqs.~\eqref{eq:conditionA} and \eqref{eq:conditionB},
we get an equation in $r_\mathrm{crit}$,
\begin{equation}
    \begin{split}
        \frac{GM}{2r_\mathrm{crit}} &- \frac{(\gamma - 1)}{2} r_\mathrm{crit} q'(r_\mathrm{crit}) \\ 
        &- \frac{2(\gamma -1)}{\gamma + 1} \left[ B - q(r_\mathrm{crit}) + \frac{GM}{r_\mathrm{crit}} \right] \equiv F(r_\mathrm{crit}) = 0.
    \end{split}
\end{equation}

Now, use a numerical root finding method
to find $r_\mathrm{crit}$.
We used 
Brent's bracketing
method from Python \texttt{scipy.optimize} module \citep{2020SciPy-NMeth}.

Once we have the value of $r_\mathrm{crit}$, we get the value of $c_\mathrm{crit}$
from Eq.~\eqref{eq:conditionA}.\newline

\noindent \textbf{Step 3:} Estimate the $dv/dr$ slope at the critical point $r=r_\text{crit}$ with l'Hospital's rule:\newline

At $r = r_\mathrm{crit}$, both the numerator and denominator of Eq.~\eqref{eq:vel_ode_apend} vanishes. 
So we need to use l'Hospital's rule to evaluate $dv/dr = g/f$, where $g$ is the terms inside the square brackets and $f = (v^2 - c_s^2)/v$.

After some manipulation, we get a quadratic equation in $\left.\frac{dv}{dr} \right|_{r_\mathrm{crit}} \equiv s$ (the steps are detailed in Appendix \ref{sec_ap:dvdr_estimation}):
\begin{equation}\label{eq:dvdr_quadratic_final}
    (\gamma + 1) s^2 + \left[ \frac{(\gamma-1)}{c_\mathrm{crit}} \left( q'(r_\mathrm{crit}) + \frac{GM}{r_\mathrm{crit}^2} \right) + \frac{2(\gamma-1)c_\mathrm{crit}}{r_\mathrm{crit}}\right] s - \left. \frac{\partial g}{\partial r} \right|_{r_\mathrm{crit}} = 0.
\end{equation}
Among the two roots of this quadratic equation $s_{\pm}$, the positive root $s_+$ gives us a physically meaningful transonic solution which accelerates outwards.
\newline

\noindent \textbf{Step 4:} Integrate for $v$ starting from $r_\mathrm{crit}$ in both directions:\newline

Now we have $\left. \frac{dv}{dr} \right|_{r_\mathrm{crit}} \equiv s_+$, $r_\mathrm{crit}$, and $c_\mathrm{crit}$. We can numerically integrate for $v$ in both $r$ directions from just outside the critical point $r_\mathrm{crit}$.

We can use, for instance, a linear Taylor step $\delta r$,
\begin{equation}
    v (r_\mathrm{crit} + \delta r) \approx c_\mathrm{crit} + s_+ \delta r,
\end{equation}
and use the Runge-Kutta (RK4) method to recover $v(r)$. At each step $r_i \rightarrow r_{i+1}$, $c_s^2(r_i, v_i)$ is evaluated.
\newline

\noindent \textbf{Step 5:} Recover other quantities:\newline

Once we have $v(r)$ in the entire domain, all other relevant physical quantities, $c_s$, $T$, $\rho$, $p$, can be obtained as a function of the radius.

\section{Estimation of \texorpdfstring{$dv/dr$}{dv/dr} at the critical radius}
\label{sec_ap:dvdr_estimation}

In Eq. \eqref{eq:vel_ode_apend},
\begin{equation}
    \frac{dv}{dr} = \frac{g(r,v)}{f(r,v)},
\end{equation}
with
\begin{equation}
    g = \frac{2c_s^2}{r} - \frac{GM}{r^2} + (\gamma-1)q'(r)
\end{equation}
and
\begin{equation}
\label{eq:f}
    f = \frac{v^2- c^2}{v}.
\end{equation}

At $r=r_\text{crit}, v = c_\text{crit}$, which gives $f = 0,~g=0$.
We apply l'Hospital's rule,
\begin{equation}\label{eq:lhospital_condition}
    s\equiv \frac{dv}{dr} = \frac{\frac{d}{dr}g|_{r_\text{crit}}}{\frac{d}{dr}f|_{r_\text{crit}}},
\end{equation}
where
\begin{equation}
    \frac{df}{dr} = \frac{\partial f}{\partial r} + \frac{\partial f}{\partial v}
    \frac{dv}{dr}
\end{equation}
and
\begin{equation}
        \frac{\partial f}{\partial r} = -\frac{1}{v} \frac{\partial c_s^2}{\partial r}
                                        =\frac{(\gamma-1)}{c_\text{crit}}\left(q'(r) + \frac{GM}{r^2}\right),
\end{equation}
from the Bernoulli integral. 
Eq.~\eqref{eq:f} gives
\begin{equation}
        \frac{\partial f}{\partial v} =1 + \frac{c_s^2}{v^2} - \frac{1}{v}\frac{\partial c_s^2}{\partial v} 
                                        = 1 + \frac{c_s^2}{v^2} + (\gamma -1).
\end{equation}
So, at $v = c_\text{crit}$,
\begin{equation}
    \left.\frac{\partial f}{\partial v}\right|_{r_\text{crit}} = \gamma + 1.
\end{equation}
So, we have the total derivative of $f$ as,
\begin{equation}
   \left.\frac{df}{dr}\right|_{r_\text{crit}} 
   = \frac{\gamma -1}{c_\text{crit}}\left(q'(r) + \frac{GM}{r_\text{crit}^2}\right)  + (\gamma + 1)\frac{dv}{dr}.
\end{equation}

In a similar way,
\begin{equation}
    \frac{dg}{dr} = \frac{\partial g}{\partial r} + \frac{\partial g}{\partial v} 
    \frac{dv}{dr}
\end{equation}
with
\begin{equation}
    \frac{\partial g}{\partial v} = \frac{2}{r}\frac{\partial c_s^2}{\partial v} = -\frac{2}{r} (\gamma - 1)v.
\end{equation}
At $r = r_\text{crit}$, $v = c_\text{crit}$, 
\begin{equation}
    \left.\frac{\partial g}{\partial v}\right|_{r_\text{crit}} = -\frac{2}{r_\text{crit}} (\gamma -1 )c_\text{crit}.
\end{equation}
Now,
\begin{equation}\label{eq:dogdor}
    \frac{\partial g }{\partial r} = \frac{2}{r} \frac{\partial c_s^2}{\partial r} - \frac{2c_s^2}{r^2} + \frac{2GM}{r^3} + (\gamma - 1)q''(r).
\end{equation}
At $r=r_\text{crit},~\frac{\partial c_s^2}{\partial r} = (\gamma-1)\left[-q'(r_\text{crit}) - \frac{GM}{r_\text{crit}^2}\right]$; putting this to Eq. \eqref{eq:dogdor}, we get 
\begin{equation}\label{eq:dogdor_rcrit}
\begin{split}
        \left.\frac{\partial g}{\partial r}\right|_{r_\text{crit}} = &-\frac{2}{r_\text{crit}}(\gamma-1)\left[q'(r_\text{crit}) +\frac{GM}{r_\text{crit}^2}\right] \\
    &- \frac{2c_\text{crit}^2}{r_\text{crit}^2} + \frac{2GM}{r_\text{crit}^3} + (\gamma - 1) q''(r_\text{crit}).
\end{split}
\end{equation}
But from Eq. \eqref{eq:conditionA}, we have,
\begin{equation}
    \frac{2c_\text{crit}^2}{r_\text{crit}^2} = \frac{GM}{r_\text{crit}^3} - \frac{(\gamma-1)}{r_\text{crit}}q'(r_\text{crit}).
\end{equation}
Putting this back to the Eq. \eqref{eq:dogdor_rcrit}, and by collecting terms together, we get
\begin{equation}\label{eq:dogdor_constr}
    \left.\frac{\partial g}{\partial r}\right|_{r_\text{crit}}= -\frac{(\gamma-1)}{r_\text{crit}}q'(r_\text{crit}) +  \frac{GM}{r_\text{crit}^3}(3 - 2\gamma) + (\gamma-1)q''(r).
\end{equation}
So, we have the total derivative of $g$ as,
\begin{equation}
\begin{split}
    \left.\frac{dg}{dr}\right|_{r_\text{crit}} = -\frac{\gamma-1}{r_\text{crit}}q'(r_\text{crit}) +  \frac{GM}{r_\text{crit}^3}(3 - 2\gamma) + (\gamma-1)q''(r) \\
    - \frac{2(\gamma-1)c_\text{crit}}{r_\text{crit}}  s
\end{split}
\end{equation}
Substituting for $\left.\frac{df}{dr}\right|_{r_\text{crit}}$ and $\left.\frac{dg}{dr}\right|_{r_\text{crit}}$ to Eq. \eqref{eq:lhospital_condition},
\begin{equation}
    s\left[\frac{(\gamma-1)}{c_\text{crit}}\left(q'(r_\text{crit}) + \frac{GM}{r_\text{crit}^2}\right) + (\gamma+1) s\right] = \frac{\partial g}{\partial r}|_{r_\text{crit}} - \frac{2(\gamma-1) c_\text{crit}}{r_\text{crit}} s.
\end{equation}
This gives us a quadratic equation in $s (\equiv \frac{dv}{dr})$, given in Eq. \eqref{eq:dvdr_quadratic_final}.


\comment{

-- To look for a transonic solution, first we need to compute the Bernoulli's constant $B$. For this, we can take the stellar parameters from a star marked in Fig. 1 For instance, $M_* = 1.052 ~\mathrm{M}_\odot$, $R_* = 280.2~\mathrm{R_\odot}$, $T_{eff} = 2862~\mathrm{K}$.
We can take $\gamma$ as either $1.02$ (almost isothermal) or $1.2$.

-- The centre of the heating potential ($r_0$) can be placed outside or inside the stellar radius. 

-- Different values of base velocity ($v_0$) can be tried out to check for a transonic solution. 

}



\bibliographystyle{aa} 
\bibliography{refs}

\end{document}